\documentclass[aps,prb,twocolumn,superscriptaddress,showpacs,floatfix]{revtex4}

\usepackage{graphicx}
\graphicspath{{figs/}}
\begin{document}
\title{Joule heating and thermoelectric properties in short single-walled carbon nanotubes: electron-phonon interaction effect}
\author{Jin-Wu~Jiang}
    \affiliation{Department of Physics and Centre for Computational Science and Engineering,
             National University of Singapore, Singapore 117542, Republic of Singapore }
\author{Jian-Sheng~Wang}
    \affiliation{Department of Physics and Centre for Computational Science and Engineering,
                 National University of Singapore, Singapore 117542, Republic of Singapore }

\date{\today}
\begin{abstract}
The electron-phonon interaction (EPI) effect in single-walled carbon nanotube is investigated by the nonequilibrium Green's function approach within the Born approximation. Special attention is paid to the EPI induced Joule heating phenomenon and the thermoelectric properties in both metallic armchair (10, 10) tube and semiconductor zigzag (10, 0) tube.
For Joule heat in the metallic (10, 10) tube, the theoretical results for the breakdown bias voltage is quite comparable with the experimental value. It is found that the Joule heat can be greatly enhanced by increasing the chemical potential, while the role of the temperature is not so important for Joule heat. In zigzag (10, 0) tube, the Joule heat is smaller than the armchair tube, resulting from nonzero band gap in the electron band structure.
For the electronic conductance $G_{e}$ and electron thermal conductance $\sigma_{el}$, the EPI has important effect at higher temperature or higher chemical potential. Compared with ballistic transport, there is an opposite tendency for $G_{e}$ to decrease with increasing temperature after EPI is considered. This is due to the dominant effect of the electron phonon scattering mechanism in the electron transport in this situation. There is an interesting `electron-drag' phenomenon for the phonon thermal conductance in case of low temperature and high chemical potential, where phonons are dragged by electrons from low temperature region into high temperature region through EPI effect.
\end{abstract}

\pacs{73.63.Fg, 72.10.Di, 72.20.Pa, 66.70.-f}
\maketitle

\pagebreak

\section{introduction}
The single-walled carbon nanotubes (SWCNT) were discovered in 1993\cite{BethuneDS,IijimaS}, and have attracted a lot of research interest from both theoretical and experimental communities (for review, see e.g. Ref.~\onlinecite{BaughmanRH,AvourisP,CharlierJC}). The SWCNT can be considered as a cylinder from rolling up of a two-dimensional graphene sheet in a particular direction, and it can be either metallic or semiconductor depending on the rolling direction.\cite{WhiteCT} Considering its good electronic performance, one promising application of the SWCNT is to design functional electronic nanodevice.\cite{NiuC,KongJ,JaveyA2002,JaveyA2003,WuJ,HuR} In these nanoscale electronic devices, the Joule heating due to the electron-phonon interaction (EPI) can be a disaster. This Joule heating is not only a waste of energy, but also can break down devices. There have been some studies on the in-air breakdown of metallic SWCNT by Joule heating.\cite{SeidelRV,JaveyA2004,HataK,PopE2005,TsutsuiM,PopE2007} We will visit this Joule heating topic in the SWCNT by the nonequilibrium Green's function (NEGF) approach in this paper.

Another important physical property is the thermoelectricity, which is characterized by the figure of merit ($ZT$). The pursuit of high value of $ZT$ has been in process for some decades, yet the $ZT$ of most commonly used thermoelectric materials is still on the order of 1.\cite{MahanGD1997,ChenG,NolasGS,DresselhausM} In recent years, owning to the development of the nanotechnology, some nano materials are found to exhibit high $ZT$, eg. the silicon nanowire with rough surface.\cite{HochbaumAI} As the thermoelectricity is an important property, there have been large amount of studies focusing on this topic.\cite{VoTTM,ChenX,MarkussenTPRB,MartinP,DonadioD,MarkussenT,NiX,JiangJW} In a recent experiment, it was proposed that the EPI should play an important role in the thermoelectric properties.\cite{OhtaH, ChoiWS} However, to the best of our knowledge, there is still quite few theoretical works discussing how the EPI can affect the thermoelectric properties in the SWCNT. This is another topic of this paper.

In this paper, we study the EPI effect in SWCNT by using the NEGF approach under the Born approximation. The metallic armchair SWCNT(10, 10) and semiconductor zigzag SWCNT(10, 0) are comparatively studied. We are interest in two phenomena: the Joule heating and thermoelectric properties affected by EPI. We find that in all tubes, the Joule heating is sensitive to the chemical potential applied, and only shows slightly dependence on the temperature. From the Joule heating we can obtain the in-air breakdown bias voltage $V_{BD}$. The theoretical result for $V_{BD}$ agrees with the experimental value. Compared with the metallic armchair SWCNT(10, 10), the semiconductor zigzag SWCNT(10, 0) has smaller Joule heat, due to its nonzero band gap in the electron band structure.

We also investigate the thermoelectric properties affected by the EPI. We find that the EPI effect is important for the electronic conductance $G_{e}$ and electron thermal conductance $\sigma_{el}$ at higher temperature or higher chemical potential. The $G_{e}$ increases with increasing temperature in the ballistic transport. However, after the EPI is considered, we show that the $G_{e}$ exhibits decreasing behavior with the increase of temperature, resulting from the strong electron phonon scattering. For phonon thermal conductance, we find an interesting `electron-drag' phenomenon at low temperature and high chemical potential, where the phonon can be dragged by electron from low temperature region into high temperature through the EPI effect.

The present paper is organized as follows. In Sec.~II, after a discussion of the electron and phonon Hamiltonian, we present how to consider the EPI effect in NEGF approach within Born approximation. Subsec.~III~A is devoted to the results of Joule heat. The calculation method for Joule heat is discussed. Subsec.~III~B is for the results of thermoelectric properties. The numerical methods for different physical quantities are presented. The paper ends with a brief summary in Sec~IV. 

\section{theory and formula}
\subsection{Phonon and electron Hamiltonian}
The lattice dynamic properties of the SWCNT are described by the empirical Brenner inter-atomic potential\cite{Brenner} implemented in GULP\cite{Gale}. GULP is applied to optimized the structure and generate the force constant matrix which is needed in the phonon Green's function. The electron Hamiltonian is obtained in the single $\pi$ orbital tight-binding approximation scheme\cite{MahanGD2003}:
\begin{eqnarray}
H_{0} & = & J_{0}\sum_{j\delta}\left[C_{A,j}^{\dagger}C_{B,j+\delta}+C_{B,j+\delta}^{\dagger}C_{A,j}\right],\\
H_{{\rm epi}} & = & J_{1}\sum_{j\delta}\left[C_{A,j}^{\dagger}C_{B,j+\delta}+C_{B,j+\delta}^{\dagger}C_{A,j}\right]\nonumber\\
&&\times\left[\left(\vec{u}_{B,j+\delta}-\vec{u}_{A,j}\right)\cdot\hat{e}_{j,\delta}\right],\\
\hat{e}_{j,\delta} & = & \frac{\vec{r}_{B,j+\delta}-\vec{r}_{A,j}}{|\vec{r}_{B,j+\delta}-\vec{r}_{A,j}|},
\end{eqnarray}
where $J_{0} = 3.0$ eV and $J_{1} = -6.0$ eV/{\rm \AA} are the onsite potential and hoping parameter. $A$ and $B$ are atoms in the two sublattices of SWCNT. $\hat{e}_{j,\delta}$ is the direction from atom $(j,A)$ to its neighboring atom $(j+\delta,B)$. The hoping parameter $J_{1}$ is a negative value, which indicates that it is more difficult for electrons to hop between two atoms if the distance between them increases. After some algebra, we can obtain a general form for the EPI Hamiltonian of a system with $N$ atoms:
\begin{eqnarray}
\vec{M}_{lm}^{n} & = & (\delta_{nm}-\delta_{nl}) J_{1}\hat{e}_{l,m},\\
\hat{e}_{l,m} & = & \frac{\vec{r}_{m}-\vec{r}_{l}}{|\vec{r}_{m}-\vec{r}_{l}|},\\
H_{\rm epi} & = & \sum_{l,m=1}^{N}\sum_{n=1}^{N}\vec{M}_{lm}^{n}\cdot C_{l}^{\dagger}C_{m}\vec{u}_{n}\nonumber\\
&=&\sum_{l=1}^{N}\sum_{m}^{FNN}\sum_{n=l,m}\vec{M}_{lm}^{n}\cdot C_{l}^{\dagger}C_{m}\vec{u}_{n},
\end{eqnarray}
where the summation of $m$ is taken over the first-nearest-neighbors of atom $l$. And the third index, $n$, can either be atom $l$ or $m$.
\begin{figure}[htpb]
  \begin{center}
    \scalebox{1.2}[1.2]{\includegraphics[width=7cm]{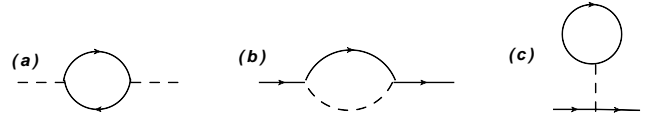}}
  \end{center}
  \caption{(Color online) Feynman diagram for EPI in the Born approximation. (a) is the phonon self-energy; (b) and (c) are the electron self-energy.}
  \label{fig_feynman}
\end{figure}
In this general formula, the choice of atom $l$ is arbitrary, not restricted to sublattice $A$ anymore. It is straightforward to show the following two symmetries of the $M$ matrix:
\begin{eqnarray}
\vec{M}_{lm}^{n} & = & \vec{M}_{ml}^{n},\\
\vec{M}_{lm}^{l} & = & -\vec{M}_{lm}^{m}.
\end{eqnarray}
We note that $\vec{M}_{lm}^{n}$ is a vector with the upper script $n$ as an atom index. It can be written in another form $M_{lm}^{n,\alpha}$, where $\alpha=x, y, z$. Or we will simply use notation $M_{lm}^{n}$ with $n$ as the $n$-th degree of freedom.

\subsection{EPI effect under the Born approximation}
In the Born approximation, the phonon/electron self-energy are expanded in terms of the EPI Hamiltonian, and the expansion is truncated up to the second order.\cite{HaugH,LuJT,WangJS2008} Fig.~\ref{fig_feynman} are the three corresponding Feynman diagrams for the phonon/electron self-energy. The dot/solid lines corresponds to the phonon/electron Green's functions without EPI. If these non-perturbed Green's functions are replaced by full Green's functions with EPI, then the full Green's functions can be obtained self-consistently. This is the self-consistent Born approximation.\cite{HaugH} It is a big challenge to apply the self-consistent Born approximation in a real system like SWCNT in this paper, since the self-consistent solution is very difficult to achieve for system with lots of degrees of freedom. There are also some other approximations to study the EPI effect. For example, in the lowest order expansion approximation,\cite{ViljasJK,PaulssonM,FrederiksenT} it is assumed that the single-particle Green's functions and lead self-energies are independent of the energy. This assumption can simplify the integrals over energy, while retains the Pauli exclusion principle for fermionic particles.

We apply the Born approximation so as to get valid results within reasonable calculation cost for SWCNT. Fig.~\ref{fig_feynman}~(a) is the EPI induced phonon self-energy. The lesser and retarded versions of this self-energy will be of use:
\begin{eqnarray}
&&\Pi_{nq}^{<}[\omega]  =  \left(-i\right)\sum_{lm}\sum_{op}M_{lm}^{n}M_{op}^{q}\frac{1}{2\pi}\nonumber\\
&&\int_{-\infty}^{+\infty}G_{0,mo}^{<}[\epsilon]G_{0,pl}^{>}[\epsilon-\hbar\omega]d\epsilon,\\
&&\Pi_{nq}^{r}[\omega] = \left(-i\right)\sum_{lm}\sum_{op}M_{lm}^{n}M_{op}^{q}\frac{1}{2\pi}\times\nonumber\\
&&\int_{-\infty}^{+\infty}\left(G_{0,mo}^{r}[\epsilon]G_{0,pl}^{<}[\epsilon-\hbar\omega]+G_{0,mo}^{<}[\epsilon]G_{0,pl}^{a}[\epsilon-\hbar\omega]\right)d\epsilon.\nonumber\\
\end{eqnarray}
Diagram (b) is the EPI induced electron self-energy. Its lesser and retarded versions are:
\begin{eqnarray}
&&\Sigma_{lp}^{<}[\epsilon] = i\hbar\sum_{mn}\sum_{oq}M_{lm}^{n}M_{op}^{q}\frac{1}{2\pi}\nonumber\\
&&\int_{-\infty}^{+\infty}G_{0,mo}^{<}[\epsilon-\hbar\omega]D_{0,nq}^{<}[\omega]d\omega,\\
&&\Sigma_{lp}^{r,1}[\epsilon] = i\hbar\sum_{mn}\sum_{oq}M_{lm}^{n}M_{op}^{q}\frac{1}{2\pi}\nonumber\\
&&\int_{-\infty}^{+\infty}\big(G_{0,mo}^{r}[\epsilon-\hbar\omega]D_{0,nq}^{r}[\omega]\nonumber\\
&&+G_{0,mo}^{r}[\epsilon-\hbar\omega]D_{0,nq}^{<}[\omega]
 +G_{0,mo}^{<}[\epsilon-\hbar\omega]D_{0,nq}^{r}[\omega]\big)d\omega,\nonumber\\
\end{eqnarray}
Diagram (c) is another EPI induced self-energy for the electron. Resulting from the loop in the diagram, the lesser version self-energy is zero. The retarded version for this self-energy is:
\begin{eqnarray}
\Sigma_{op}^{r,2}[\epsilon] &=& \left(-i\right)\frac{1}{2\pi}\sum_{lmn}\sum_{q}M_{lmn}M_{opq}\nonumber\\
&\big[&\int_{-\infty}^{+\infty}\left(G_{0,ml}^{r}[\epsilon]+G_{0,ml}^{<}[\epsilon]\right)d\epsilon D_{0,qn}^{r}[0]\nonumber\\
 &+&\int_{-\infty}^{+\infty}\left(G_{0,ml}^{r}[\epsilon]+G_{0,ml}^{a}[\epsilon]\right)d\epsilon D_{0,qn}^{<}[0]\big].
\end{eqnarray}
We can see this self-energy is actually a constant matrix. Its effect is simply to shift the electron energy level by a constant.

After the EPI induced self-energy are obtained as shown above, the full Green's function of the system can be calculated from Dyson equations:
\begin{eqnarray}
D^{r}\left[\omega\right] & = & \left[D_{0}^{r-1}-\Pi_{\rm epi}^{r}\right]^{-1},\\
D^{a}\left[\omega\right] & = & \left(D^{r}\left[\omega\right]\right)^{\dagger},\\
D^{<}\left[\omega\right] & = & D^{r}\left[\omega\right] \Pi^{<}[\omega] D^{a}\left[\omega\right],\\
G^{r}\left[\epsilon\right] & = & \left[G_{0}^{r-1}-\Sigma_{\rm epi}^{r}\right]^{-1},\\
G^{a}\left[\epsilon\right] & = & \left(G^{r}\left(\epsilon\right)\right)^{\dagger},\\
G^{<}\left[\epsilon\right] & = & G^{r}\left[\epsilon\right] \Sigma^{<}[\epsilon] G^{a}\left[\epsilon\right],
\end{eqnarray}
where $\Pi^{<}=\Pi_{L}^{<}+\Pi_{R}^{<}+\Pi^{<}_{epi}$ and $\Sigma^{<}=\Sigma_{L}^{<}+\Sigma_{R}^{<}+\Sigma^{<}_{epi}$ are the total phonon/electron self-energy.

The thermal or electronic current from the leads into center can be formulated through the full Green's function.\cite{LuJT,WangJS2008,HaugH} The phonon thermal current flowing from left lead into the system is:
\begin{eqnarray}
I_{L}^{ph}&=&-\frac{1}{2\pi}\int_{-\infty}^{+\infty}d\omega \hbar\omega {\rm Tr}(D^r[\omega]\Pi_{L}^{<}[\omega] + D^<[\omega]\Pi_{L}^{a}[\omega]),\nonumber\\
\label{eq_Iph}
\end{eqnarray}
where $\Pi_{L}$ is the self-energy due to the coupling of the system to the left lead. The right lead is analogous.

In parallel, the electronic current from left lead is:
\begin{eqnarray}
J_{L}=e\frac{1}{2\pi}\int_{-\infty}^{+\infty}d\epsilon {\rm Tr}\left(G^r[\epsilon]\Sigma_{L}^{<}[\epsilon] + G^<[\epsilon]\Sigma_{L}^{a}[\epsilon]\right).
\label{eq_J}
\end{eqnarray}
The electron thermal current from left lead into the system is:
\begin{eqnarray}
I_{L}^{el}&=&\frac{-1}{2\pi}\int_{-\infty}^{+\infty}d\epsilon (\epsilon-\mu _{L}) {\rm Tr}\left(G^r[\epsilon]\Sigma_{L}^{<}[\epsilon] + G^<[\epsilon]\Sigma_{L}^{a}[\epsilon]\right),\nonumber\\
\label{eq_Iel}
\end{eqnarray}
where $\mu_{L}$ is the chemical potential of the left lead.

\section{results and discussion}
\subsection{Joule heating}
In systems without EPI, the phonon thermal conductance is purely due to the temperature difference of the left and right leads. In this situation, it can be shown that the phonon thermal current from two leads has the same magnitude while opposite in direction\cite{Zeng}; i.e $I_{L}^{ph}=-I_{R}^{ph}$. This simply means that the phonon thermal current runs out of one lead and jump into the other lead. We refer to this kind of thermal current as normal thermal current. In a system with EPI, the phonon thermal current can also be generated by applying temperature difference in left and right leads.
\begin{figure}[htpb]
  \begin{center}
    \scalebox{1.2}[1.2]{\includegraphics[width=7cm]{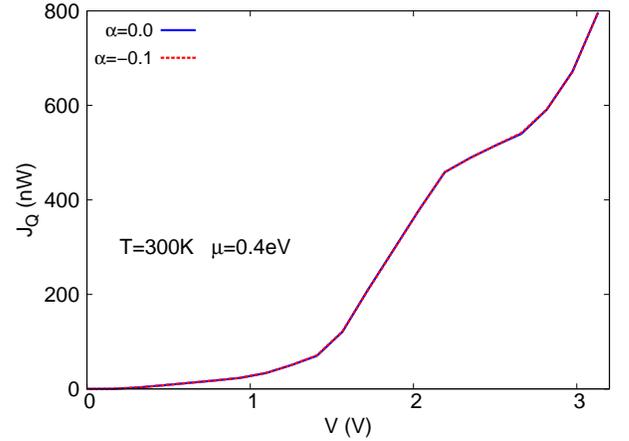}}
  \end{center}
  \caption{(Color online) SWCNT(10, 10). Joule heat calculated with different parameters of $\alpha$. $\Delta T=2\alpha T$.}
  \label{fig_t_a_J_alpha}
\end{figure}
 However, besides this normal effect there is another method to induce the phonon thermal current. It can be driven by introducing bias voltage without temperature gradient across the system. In this situation, phonons are excited through scattering with moving electrons. This is the microscopic mechanism of the Joule heating. The Joule heat will transfer from the system into two leads. The total Joule heat current is the collection of these heat currents: $J_{Q}=I_{L}^{ph}+I_{R}^{ph}$. We point out that this formula still holds even if there is normal heat current due to temperature difference between two leads. Because the normal thermal current has the relation $I_{L}^{ph}=-I_{R}^{ph}$, as a result the normal heat current from left and right leads cancel with each other in the formula of $J_{Q}$. This issue will be further discussed in the following.

\subsubsection{armchair SWCNT (10, 10)}
Fig.~\ref{fig_t_a_J_alpha} shows the Joule heat $J_{Q}$ at different bias voltage $V$ with $T=300$ K and $\mu=0.4$ eV. The Fermi energy $\epsilon_{F}=0$. The length of the SWCNT here is 1.5 nm with 240 carbon atoms. In the experiment, it is advantageous to prepare SWCNT samples with length in micrometer scale; while the quantum mechanical NEGF theoretical approach prefers shorter SWCNT due to large memory required.
\begin{figure}[htpb]
  \begin{center}
    \scalebox{1.0}[1.0]{\includegraphics[width=7cm]{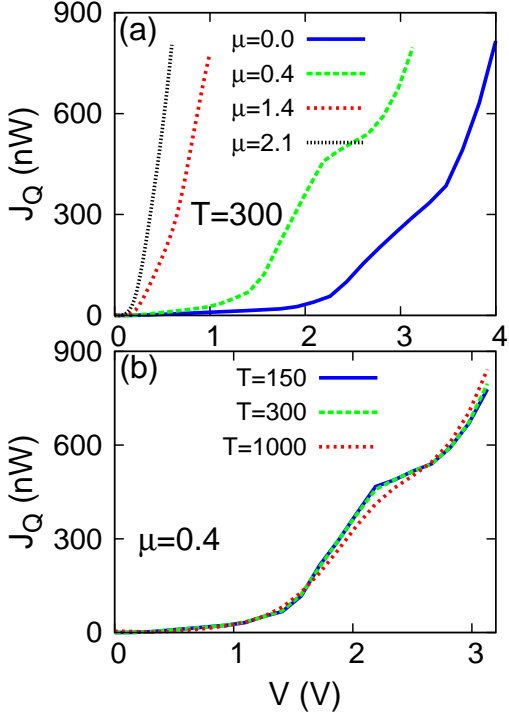}}
  \end{center}
  \caption{(Color online) SWCNT(10,10). Joule heating V.S. $V$ at different chemical potential (a), and at different temperature (b).}
  \label{fig_t_a_J_V}
\end{figure}
\begin{figure}[htpb]
  \begin{center}
    \scalebox{1.0}[1.0]{\includegraphics[width=7cm]{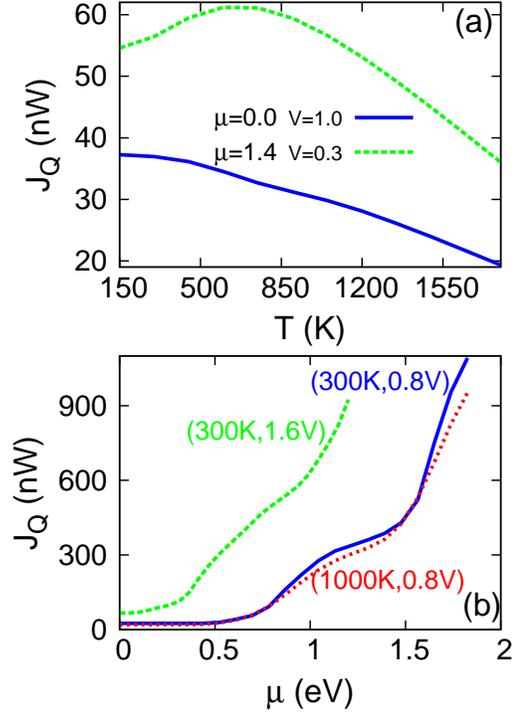}}
  \end{center}
  \caption{(Color online) SWCNT(10,10). Joule heat with different chemical potential or temperature. (a). Joule heating V.S. $T$. (b). Joule heating V.S. $\mu$.}
  \label{fig_t_a_J_T_u}
\end{figure}
 So we focus on short SWCNT with length in nanometer scale. The parameter $\alpha$ is the temperature difference ratio, i.e $T_{R}-T_{L}=2\alpha T$. It shows that the value of Joule heat does not depend on whether there is temperature difference between two leads or not. As we have mentioned in the above, this is because the normal phonon thermal transport due to temperature difference of the left and right leads cancels with each other in the formula $J_{Q}=J_{L}+J_{R}$. In the rest of the text, if we calculate the Joule heat, without mentioning it, we set a constant temperature for the system, i.e $\alpha=0$. In this case, the phonon thermal current calculated from Eq.~(\ref{eq_Iph}) are purely the Joule heat. We note that in some figures of this manuscript, the quantity units are omitted. They are in terms of SI units: [$T$]=K, [$\mu$]=eV, and [$V$]=V.

Fig.~\ref{fig_t_a_J_V} is the Joule heat V.S. bias voltage $V$. Panel (a) shows the result for a constant temperature 300K. If the chemical potential $\mu$ is zero or very small, the $J_{Q}$ increases very slowly with increasing bias voltage $V$, and the value of $J_{Q}$ is obvious only after $V>2$ V. If the chemical potential $\mu$ is considerable large, eg. 1.4 eV, or 2.1 eV, the Fermi surface goes into the second/third energy level of the electron conduction band. As a result, the conductive electron density is greatly enhanced and much more phonons will be excited through EPI, leading to rapid increase of Joule heat with increasing $V$. A large value of Joule heat can be achieved at bias voltage $V=0.5$ V. These theoretical results indicate that in short SWCNT with length $L$ in the nanoscale, the electron transport is almost in ballistic region even in the existence of EPI. So the electron carrier density need to be greatly enhanced so as to generate remarkably large Joule heat. Panel (b) shows the Joule heat V.S. $V$ at different temperatures. Only limited influence from the temperature can be found, since the energy level of the electron is in the order of eV. The temperature need to be in the order of $10^{4}$ K to introduce considerable influence on the electron doping.

Although the temperature has very small effect on the Joule heat, it is still interesting to investigate this effect in more detail. It carries some valuable information of temperature dependence for the EPI effect. Fig.~\ref{fig_t_a_J_T_u}~(a) shows the Joule heat at different temperatures. If $\mu=0$ and $V=1.0$ V, the Joule heat is monotonically decreasing function of temperature. At $T$=1800 K, the value of Joule heat is reduced by 50$\%$ compared with that at 150 K. At $\mu=1.4$ eV and $V=0.3$ V, we can observe more abundant behaviors of the Joule heat in different temperature regions. Below 600 K, the Joule heat increases with increasing temperature; then reaches a maximum value around 600 K; and finally decreases with further increasing temperature.

\begin{figure}[htpb]
  \begin{center}
    \scalebox{1.2}[1.2]{\includegraphics[width=7cm]{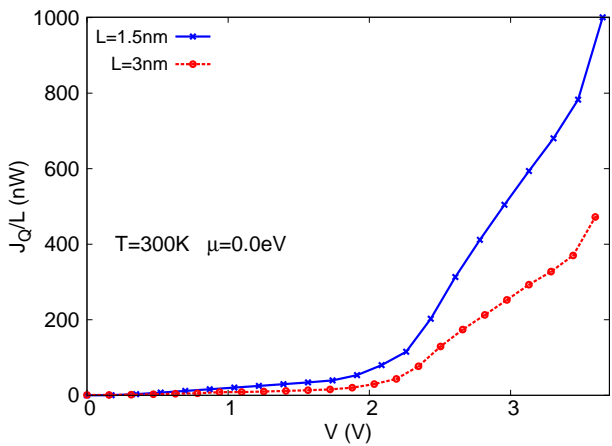}}
  \end{center}
  \caption{(Color online) SWCNT(10, 10). Joule haet per length in SWCNT with different lengths.}
  \label{fig_t_a_J_length}
\end{figure}

Fig.~\ref{fig_t_a_J_length} compares the Joule heat per length in SWCNT (10, 10) with different length $L$ at $\mu=0$ and $T=300$ K. As we can see the Joule heat per length is smaller in longer tubes, since the electron is almost in the ballistic transport region. However, if we check the Joule heat itself, we find that the Joule heat is larger in the longer tube than that in the shorter one. They are not exactly equal to each other. Especially in higher bias voltage region. The underlying mechanism is that electrons in longer tube spend more time in the system during its travel from one lead into the other. As a result, more phonon will be excited, leading to larger value of Joule heat.

Now we do some comparison of our theoretical results with experiments. In experiments, the metallic SWCNT can be put on insulating substrate, and applied a bias voltage on it.\cite{JaveyA2004} After the bias voltage is increased to a certain value, the SWCNT will breakdown and this breakdown bias voltage ($V_{BD}$) can be recorded through the I-V curve. The experimental value of $V_{BD}$ is about 2.6 V for SWCNT samples with length $10\pm 5$ nm, and diameter distributed in the range of [1.5, 2.5] nm. In our theoretical simulation, the SWCNT (10, 10) has the length as 3.0 nm and diameter 1.4 nm. They are not exactly in the same size of experimental samples. However we would like to do some qualitative comparison between our theoretical results and the experimental data. First of all, the in-air breakdown temperature for SWCNT is about $T_{BD}=873$ K, which is due to Joule heating by applying bias voltage. This value of $T_{BD}$ is obtained by the thermogravimetric analysis experiments.\cite{ChiangIW1,ChiangIW2}. In addition, the breakdown of SWCNT is taken place in its middle region by Joule heating, and the temperature in this middle region can be obtained from the heat conduction equation\cite{PopE2007}:
\begin{eqnarray}
T=T_{0}+P'/g,
\end{eqnarray}
where $T_{0}$ is the enviroment temperature, and $g\approx 0.17 $ WK$^{-1}$m$^{-1}$ is the net heat loss rate to the substrate per unit length.
\begin{figure}[htpb]
  \begin{center}
    \scalebox{1.0}[1.0]{\includegraphics[width=7cm]{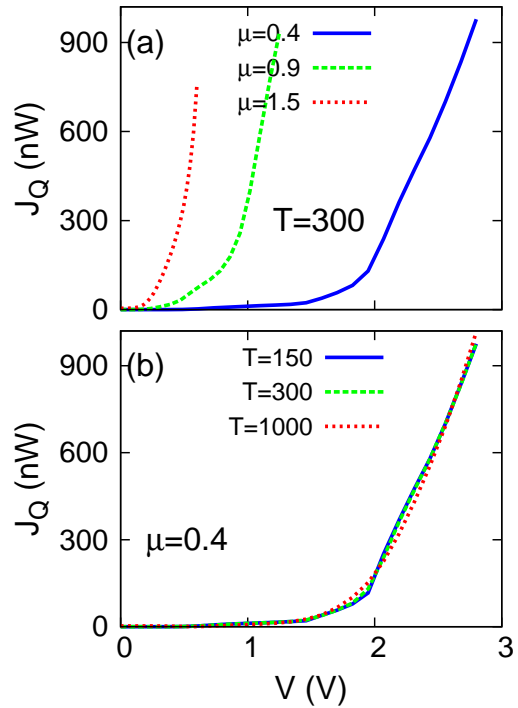}}
  \end{center}
  \caption{(Color online) SWCNT(10, 0). Joule heat V.S. bias voltage $V$ with (a) different chemical potential, and (b)different temperature.}
  \label{fig_t_z_J_V}
\end{figure}
 $P'$ is the Joule heat per unit length. So the breakdown Joule heat per unit length can be obtained from this formula: $P'_{BD}=(T_{BD}-T_{0})*g\approx 80$ Wm$^{-1}$. Corresponding to this value in Fig.~\ref{fig_t_a_J_length}, we can obtain the theoretical breakdown bias voltage $V_{BD}\approx 2.5$ V, which is quite comparable with the experiment.

\subsubsection{zigzag SWCNT (10, 0)}
\begin{figure}[htpb]
  \begin{center}
    \scalebox{1.2}[1.2]{\includegraphics[width=7cm]{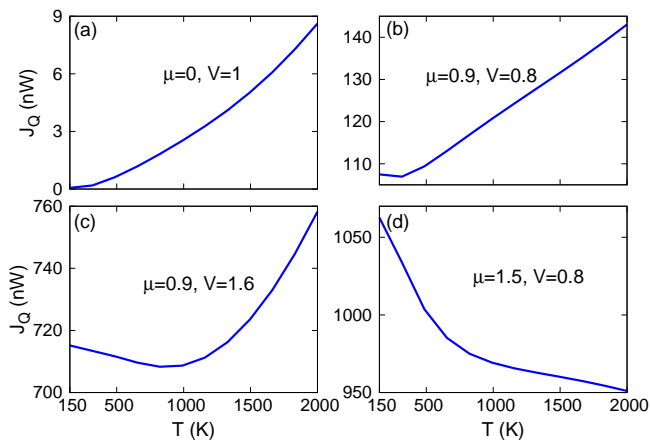}}
  \end{center}
  \caption{(Color online) SWCNT(10, 0). Joule heat V.S. temperature at different fixed chemical potential.}
  \label{fig_t_z_J_T}
\end{figure}

Fig.~\ref{fig_t_z_J_V} is the Joule heat V.S. $V$ in the zigzag SWCNT(10, 0). It has a similar behavior as the armchair SWCNT(10, 10) shown in Fig.~\ref{fig_t_a_J_V}. In case of small $\mu$, the Joule heat keeps small upto $V=1.0$ V. If chemical potential is larger, the Joule heat increases quickly with the increase of bias voltage $V$. The temperature only plays a weak role on the Joule heat. Besides these similar behaviors, we can see some interesting difference between these two types of SWCNT. The SWCNT(10, 0) has a band gap of about 1.2 eV in its electron band structure, while SWCNT(10, 10) is gapless. So the former is semiconducting while the later is a metallic tube. As a result, in small $\mu$ region the Joule heat in the zigzag SWCNT(10, 0) should be much smaller than that in the armchair SWCNT(10, 10). This is confirmed in our calculation shown in Fig.~\ref{fig_t_z_J_T}~(a) and Fig.~\ref{fig_t_a_J_T_u}~(a) with $\mu=0$ and $V=1.0$ V. In whole temperature range, the Joule heat in zigzag SWCNT(10, 0) is smaller than the armchair SWCNT(10, 10) at least by a factor of four in magnitude.

Fig.~\ref{fig_t_z_J_T}~(a) also shows that the Joule heat is almost zero at low temperatures in zigzag SWCNT, resulting from the nonzero electron band gap. The Joule heat increases monotonically with increasing temperature, which is quite different from the armchair SWCNT shown in Fig.~\ref{fig_t_z_J_T}~(a) with $\mu=0$ and $V=1.0$ V. As shown in panels (b) and (c), if the $\mu$ is a small nonzero value, the Joule heat shows a small decreasing behavior in low temperature region;
\begin{figure}[htpb]
  \begin{center}
    \scalebox{1.2}[1.2]{\includegraphics[width=7cm]{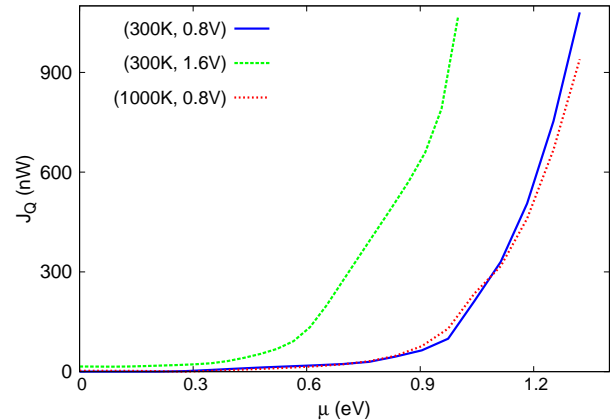}}
  \end{center}
  \caption{(Color online) SWCNT(10, 0). Joule heat V.S. chemical potential with different fixed chemical potential and bias voltage.}
  \label{fig_t_z_J_u}
\end{figure}
 then reaches a minimum value and increases with further increase of temperature. Panel (d) displays the result for large chemical potential, where the Joule heat decreases monotonically with the increase of temperature.

Fig.~\ref{fig_t_z_J_u} shows that the chemical potential also has important effect on the Joule heat in the zigzag SWCNT. If the chemical potential is high, a large Joule heat can be obtained with quite small bias voltage.

\subsection{thermoelectric properties}
The thermoelectric physical quantities can be calculated by definition from the thermal and electronic currents. This numerical calculation procedure is different from the ballistic transport where Landauer formula can be used for the thermal and electronic current. Actually, the expression in Eq.~(\ref{eq_Iph})-(\ref{eq_Iel}) can be reformulated into an effective Landauer formula after an effective transmission function is introduced.\cite{WangJS2008} However, this effective transmission function is quite different from its counterpart in the ballistic transport, because it depends on temperature (for phonon and electron) and also chemical potential (for electron). As a result, we need to do calculation numerically for the first derivative of the thermal or electronic currents with respect to the temperature or chemical potential. These first derivatives of the thermal/electronic currents are required in the calculation of different thermoelectric physical quantities. In ballistic transport, they can be simply obtained by taking first derivatives of the Bose-Einstein or Fermi-Dirac distribution functions.

\begin{figure}[htpb]
  \begin{center}
    \scalebox{1.2}[1.2]{\includegraphics[width=7cm]{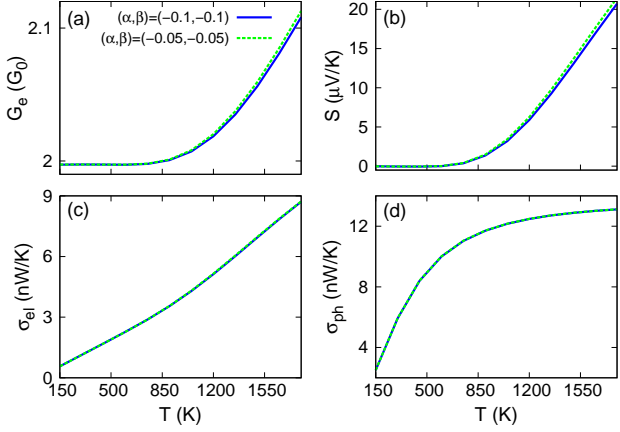}}
  \end{center}
  \caption{(Color online) SWCNT(10, 10). Thermoelectric properties calculated with different parameters $(\alpha, \beta)$. $\alpha$ and $\beta$ are the ratio of difference for the temperature and chemical potential, respectively.}
  \label{fig_t_a_zt_alpha}
\end{figure}
The Taylor expansion of the thermal and electronic currents in terms of $\Delta T=T_{R}-T_{L}$ and $\Delta\mu=\mu_{R}-\mu_{L}$ are
\begin{eqnarray}
I^{ph} & = & \frac{\partial I^{ph}}{\partial T}\Delta T+\frac{\partial I^{ph}}{\partial\mu}\Delta\mu,\\
I^{el} & = & \frac{\partial I^{el}}{\partial T}\Delta T+\frac{\partial I^{el}}{\partial\mu}\Delta\mu,\\
J & = & \frac{\partial J}{\partial T}\Delta T+\frac{\partial J}{\partial\mu}\Delta\mu,
\end{eqnarray}
where the partial differential coefficient can be obtained numerically, for example,
\begin{eqnarray}
\frac{\partial I^{ph}}{\partial T} = \frac{I^{ph}(T,\Delta T,\mu,\Delta\mu=0)}{\Delta T}.
\end{eqnarray}
Using these numerically obtained coefficients, the thermoelectric quantities can be calculated by definition.
The phonon thermal conductance is:
\begin{eqnarray}
\sigma_{ph} & = & -\frac{I^{ph}}{\Delta T}|_{\Delta \mu=0}.
\end{eqnarray}
\begin{figure}[htpb]
  \begin{center}
    \scalebox{1.2}[1.2]{\includegraphics[width=7cm]{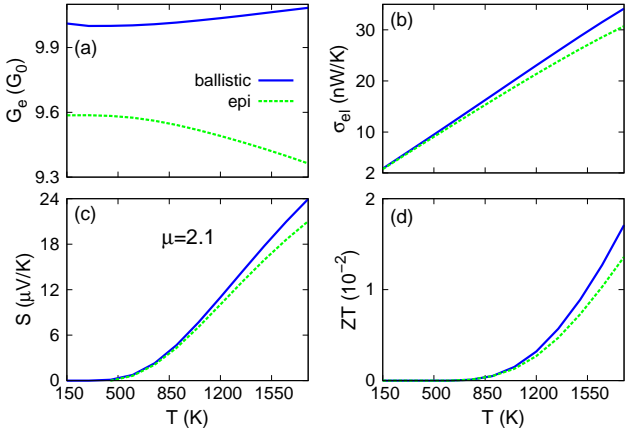}}
  \end{center}
  \caption{(Color online) SWCNT(10, 10). The EPI effect on the thermoelectric properties at fixed temperature.}
  \label{fig_t_a_zt_T}
\end{figure}
The electronic conductance is:
\begin{eqnarray}
G_{e}  =  -\frac{J}{V}|_{\Delta T=0}
  =  -e\times\frac{\partial J}{\partial\mu}.
\end{eqnarray}
The Seebeck coefficient is:
\begin{eqnarray}
S  =  -\frac{V}{\Delta T}|_{J=0}
  =  -\frac{1}{e}\times\frac{\Delta\mu}{\Delta T}|_{J=0}
  =  \frac{1}{e}\times\left(\frac{\partial J}{\partial T}/\frac{\partial J}{\partial\mu}\right).
\end{eqnarray}
The electron thermal conductance is:
\begin{eqnarray}
\sigma_{el} & = & -\frac{I^{el}}{\Delta T}|_{J=0}\nonumber\\
& = & -\frac{\frac{\partial I^{el}}{\partial T}\Delta T+\frac{\partial I^{el}}{\partial\mu}\Delta\mu}{\Delta T}|_{J=0}\nonumber\\
& = & -\frac{\partial I^{el}}{\partial T}+\frac{\partial I^{el}}{\partial\mu}\times\frac{\partial J}{\partial T}/\frac{\partial J}{\partial\mu}.
\end{eqnarray}

\subsubsection{armchair SWCNT (10, 10)}
As mentioned above, the partial differential coefficients need to be calculated numerically. A small temperature difference $\Delta T$, and chemical potential difference $\Delta \mu$ between left and right leads are introduced. The temperatures in the left and right leads are $T_{L}=(1-\alpha)T$ and $T_{R}=(1+\alpha)T$, where $T$ is the average temperature. The chemical potentials are $\mu_{L}=(1-\beta)\mu$ and $\mu_{R}=(1+\beta)\mu$ with $\mu$ as the averaged chemical potential. So $\Delta T=2\alpha T$, and $\Delta \mu=2\beta \mu$. Smaller value of $(\alpha, \beta)$ has the benefit of approaching the analytical partial differential coefficients numerically, yet too small value of $(\alpha, \beta)$ may cause larger numerical error. So a proper chosen value for $(\alpha, \beta)$ is important. Fig.~\ref{fig_t_a_zt_alpha}. compares two results for the thermoelectric quantities, with respect to $(\alpha, \beta)=(-0.1,-0.1)$ and (-0.05, -0.05). The chemical potential $\mu=0.4$ eV. It turns out that these two curves are almost indistinguishable. So $(\alpha, \beta)=(-0.1,-0.1)$ is a good choice and will be used in following calculation without special mention.

Fig.~\ref{fig_t_a_zt_T} illustrates the EPI effect on different thermoelectric properties as a function of temperature, with chemical potential $\mu=2.1$ eV. Compared with the ballistic transport, these physical quantities are generally depressed by the EPI. In low temperature region, the EPI only has small effect due to few energy and charge carriers, especially quite few phonons. With increasing temperature, more electrons will be scattered by phonons through the EPI, resulting in further reduction of these physical quantities. Panel (a) shows that the electronic conductance increases with the increase of temperature in ballistic transport,
\begin{figure}[htpb]
  \begin{center}
    \scalebox{1.0}[1.0]{\includegraphics[width=7cm]{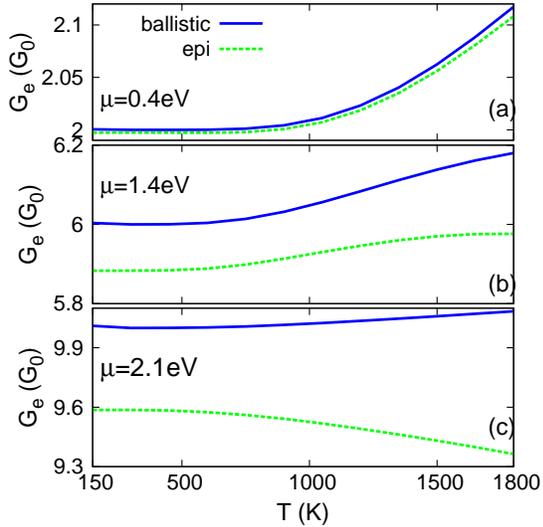}}
  \end{center}
  \caption{(Color online) SWCNT(10, 10). The electronic conductance V.S. temperature, with fixed chemical potential.}
  \label{fig_t_a_Ge_T}
\end{figure}
 but it decreases after the EPI is considered. This interesting effect will be further investigated in more detail in the following. Panel (c) and (d) exhibit that the Seebeck coefficient and $ZT$ are slightly affected by the EPI around room temperature. So in the following text we are not going to spend much discussion on these two quantities, instead we will focus on the electronic conductance $G_{e}$, electron thermal conductance $\sigma_{el}$, and the phonon thermal conductance $\sigma_{ph}$.

In Fig.~\ref{fig_t_a_Ge_T}, we do further studies on how EPI affect the $G_{e}$ with different chemical potential. There are two competing mechanisms on the $G_{e}$ with increasing temperature. On the one hand, more electrons are involved in transporting of electronic charges with the increase of temperature. It makes positive contribution to $G_{e}$. On the other hand, more phonons will also be excited at higher temperature, leading to stronger scattering of electron by phonon.
\begin{figure}[htpb]
  \begin{center}
    \scalebox{1.0}[1.0]{\includegraphics[width=7cm]{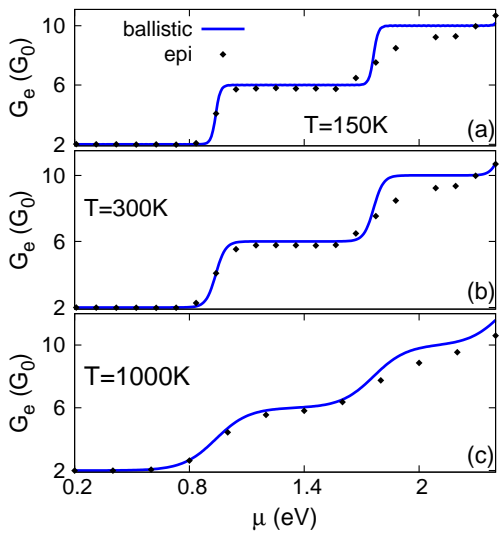}}
  \end{center}
  \caption{(Color online) SWCNT(10, 10). The electronic conductance V.S. chemical potential, at constant temperatures.}
  \label{fig_t_a_Ge_u}
\end{figure}
\begin{figure}[htpb]
  \begin{center}
    \scalebox{1.0}[1.0]{\includegraphics[width=7cm]{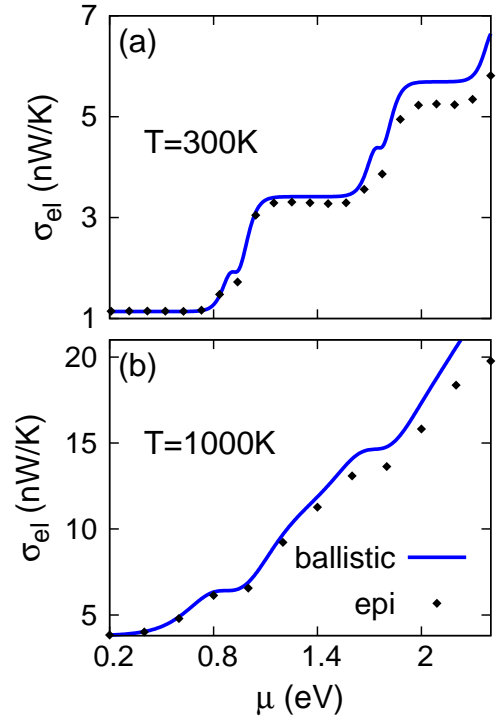}}
  \end{center}
  \caption{(Color online) SWCNT(10, 10). The electron thermal conductance V.S. chemical potential, with different fixed temperatures.}
  \label{fig_t_a_sgm_el_u}
\end{figure}
 This mechanism has negative contribution to $G_{e}$. These two competing mechanisms result in rich dynamics of the $G_{e}$. Panel (a) is for small chemical potential $\mu=0.4$ eV, which corresponds to low electron doping. In this situation, the electron number is too small to have significant electron phonon scattering. As a result, the EPI only has small effect and the positive mechanism is more important than the negative mechanism. Consequently, $G_{e}$ is slightly smaller than that in the ballistic transport, and keeps on increasing with increasing temperature. Panel (b) is for $\mu=1.4$ eV, where the EPI effect becomes more important and the negative mechanism begins to compete with the positive mechanism. As a result, $G_{e}$ exhibits obvious deviation from the ballistic result, and the increasing velocity is slowed down. Panel (c) is for large chemical potential $\mu=2.1$ eV. This is a high electron doping with Fermi level shifted into the third electron conduction band. In this case, the EPI is very important, and the negative mechanism dominants the electron transport property. So $G_{e}$ shows decreasing behavior with the increase of temperature.

Fig.~\ref{fig_t_a_Ge_u} illustrates the EPI effect on the behavior of $G_{e}$ versus $\mu$. At all studied temperatures, the EPI has important effect only in high chemical potential region. The EPI effect on the $G_{e}$ is almost ignorable
\begin{figure}[htpb]
  \begin{center}
    \scalebox{1.2}[1.2]{\includegraphics[width=7cm]{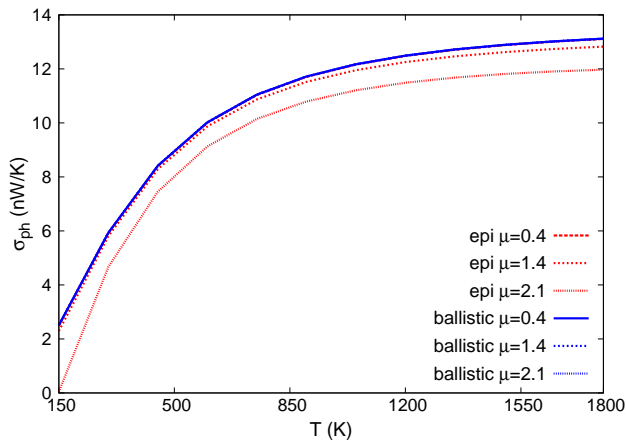}}
  \end{center}
  \caption{(Color online) SWCNT(10, 10). The phonon thermal conductance V.S. temperature, with different chemical potential.}
  \label{fig_t_a_sgm_ph_T}
\end{figure}
 in case of small chemical potential due to small number of electron. A very similar result is found for the electron thermal conductance $\sigma_{el}$ as shown in Fig.~\ref{fig_t_a_sgm_el_u}, where the EPI effect is important only if the chemical potential shifts the Fermi level into higher electron conduction band.

Fig.~\ref{fig_t_a_sgm_ph_T} shows the EPI effect on the phonon thermal conductance with different fixed chemical potential. As well known, $\sigma_{ph}$ is independent of the chemical potential in the ballistic transport. However, by considering the EPI effect, the chemical potential can affect the phonon thermal transport through electron phonon scattering.
\begin{figure}[htpb]
  \begin{center}
    \scalebox{1.0}[1.0]{\includegraphics[width=7cm]{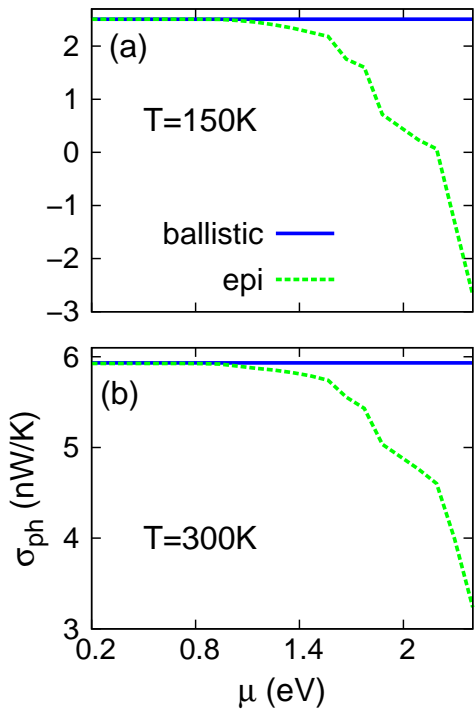}}
  \end{center}
  \caption{(Color online) SWCNT(10, 10). The phonon thermal conductance V.S. chemical potential, at constant temperatures.}
  \label{fig_t_a_sgm_ph_u}
\end{figure}
 This figure shows that $\sigma_{ph}$ decreases obviously with increasing chemical potential. Especially for high chemical potential $\mu=2.1$ eV, $\sigma_{ph}$ is almost zero in low temperature region, which means phonons are almost completely scattered by electrons. The reason is that only few phonons are excited at low temperature, yet there are huge number of conduction electrons at high chemical potential. As a result, all phonons are scattered by electrons, leading to small or zero phonon thermal conductance. At this point it is interesting enough to imagine what will happen if the chemical potential is further increased. We can anticipate that the phonon thermal conductance may become negative, which means the phonon thermal energy transfers from low temperature region into high temperature region with the help of the electron phonon scattering. Indeed, this anticipation is confirmed by our calculation shown in Fig.~\ref{fig_t_a_sgm_ph_u}. Panel (a) shows that the $\sigma_{ph}$ at 150 K decreases with increasing chemical potential, and becomes negative after $\mu>2$ eV. It is worth noting that, this effect is a kind of `electron-drag' effect and has nothing to do with the Joule heat, because in this calculation only very small bias voltage is applied $V=\Delta \mu /e=2\beta \mu /e$. The Joule heat under such small bias voltage is almost zero as can be seen from Fig.~\ref{fig_t_a_J_V}. This effect has something similar to the phonon-drag effect, which describes charge carriers dragged by phonon from hot to cold regions via momentum transfer.\cite{ScarolaVW} To see this `electron-drag' effect on phonon thermal conductance, high density of electron and low density of phonon are required; i.e high chemical potential and low temperature. Under these conditions, the SWCNT can serve as a heat pump to transfer phonon thermal energy from low temperature region into high temperature region by applying bias voltage.
\begin{figure}[htpb]
  \begin{center}
    \scalebox{1.2}[1.2]{\includegraphics[width=7cm]{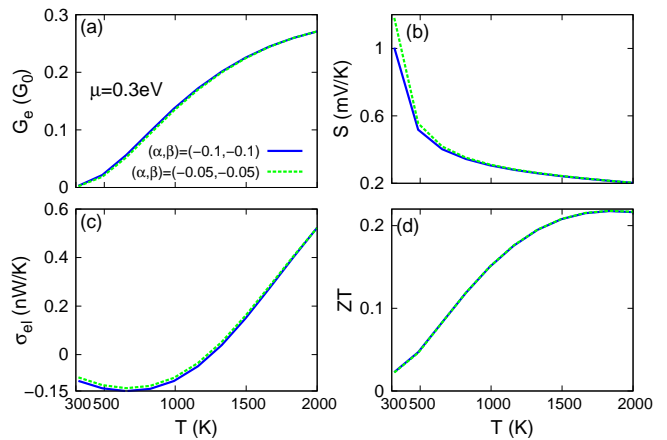}}
  \end{center}
  \caption{(Color online) SWCNT(10, 0). The thermoelectric properties calculated with different parameters ($\alpha$, $\beta$).}
  \label{fig_t_z_zt_alpha}
\end{figure}
\begin{figure}[htpb]
  \begin{center}
    \scalebox{1.0}[1.0]{\includegraphics[width=7cm]{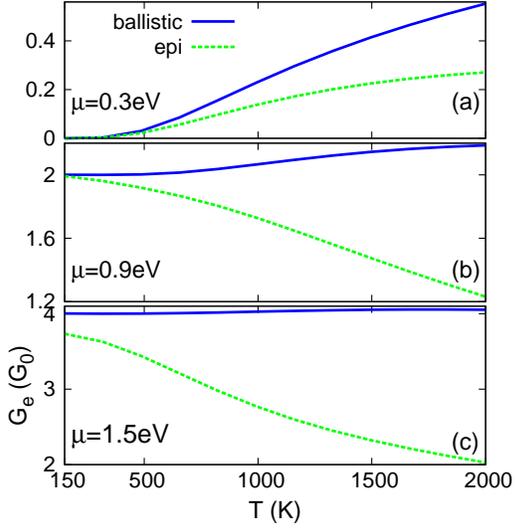}}
  \end{center}
  \caption{(Color online) SWCNT(10, 0). The electronic conductance V.S. temperature, with different fixed chemical potential.}
  \label{fig_t_z_Ge_T}
\end{figure}

\subsubsection{zigzag SWCNT (10, 0)}
The rest of the text is devoted to the thermoelectric properties in zigzag SWCNT(10, 0). Fig.~\ref{fig_t_z_zt_alpha} supports that parameters $(\alpha, \beta)=(-0.1,-0.1)$ are also suitable choice in zigzag SWCNT(10, 0). The chemical potential $\mu=0.3$ eV in the figure. We note that the Seebeck coefficient in zigzag SWCNT is 2-3 orders of magnitude larger than the armchair SWCNT, resulting from the band gap in the electron band structure.\cite{OuyangY}
\begin{figure}[htpb]
  \begin{center}
    \scalebox{0.9}[0.9]{\includegraphics[width=7cm]{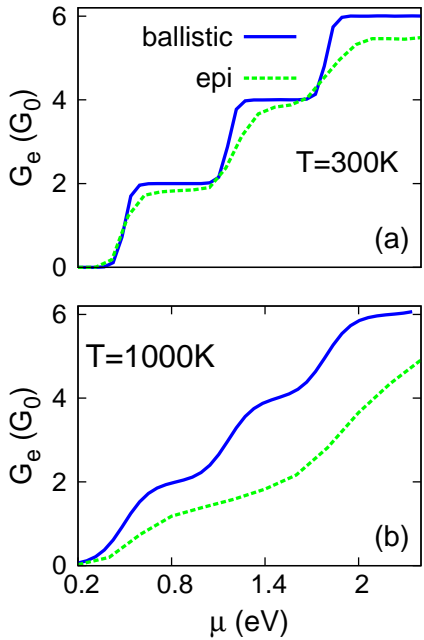}}
  \end{center}
  \caption{(Color online) SWCNT(10, 0). The electronic conductance V.S. chemical potential, at constant temperatures.}
  \label{fig_t_z_Ge_u}
\end{figure}
 For the same reason, the electron thermal conductance is much smaller in the zigzag SWCNT. As a result, the figure of merit $ZT$ is considerably larger in zigzag SWCNT.

Fig.~\ref{fig_t_z_Ge_T} is the electronic conductance $G_{e}$ in zigzag SWCNT(10,0) with different chemical potential. Compare this figure with the result of armchair SWCNT(10,10) shown in Fig.~\ref{fig_t_a_Ge_T}, some similar and different behaviors can be found. In both figures, the $G_{e}$ in ballistic transport will increase with increasing temperature. If the EPI is taken into account, the results depend on the value of chemical potential. For low chemical potential$\mu=0.3$ eV, $G_{e}$ only deviates slightly from ballistic result in low temperature region; and this deviation increases with increasing temperature, as more phonons are excited at higher temperature leading to stronger electron-phonon scattering. However, $G_{e}$ keeps on increasing as the temperature increases, which is same as ballistic transport. For high chemical potential $\mu=$ 0.9, 1.5 eV, the effect of EPI becomes more and more important, because the possibility of electron-phonon scattering increases as the chemical potential increases. So the $G_{e}$ is much smaller than the ballistic result, especially in the high temperature region, and $G_{e}$ will decrease with increasing temperature which is opposite from the ballistic case. The difference between Fig.~\ref{fig_t_z_Ge_T} and Fig.~\ref{fig_t_a_Ge_T} is quite obvious. At low chemical potential, the $G_{e}$ in SWCNT(10,0) is almost zero at low temperature, and is much smaller than that in SWCNT(10,10) in whole temperature range. This is because of the band gap in the electron band structure in the zigzag SWCNT(10,0),
\begin{figure}[htpb]
  \begin{center}
    \scalebox{1.0}[1.0]{\includegraphics[width=7cm]{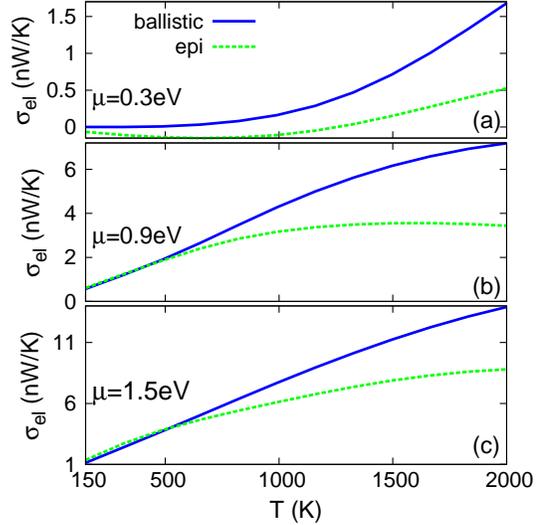}}
  \end{center}
  \caption{(Color online) SWCNT(10, 0). The electron thermal conductance V.S. temperatures, with different chemical potential.}
  \label{fig_t_z_sgm_el_T}
\end{figure}
\begin{figure}[htpb]
  \begin{center}
    \scalebox{1.2}[1.2]{\includegraphics[width=7cm]{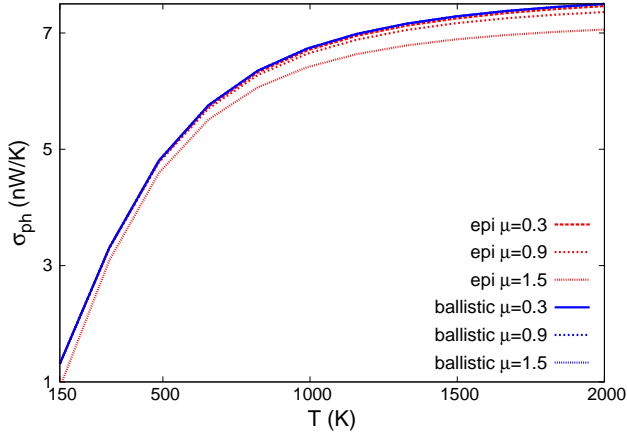}}
  \end{center}
  \caption{(Color online) SWCNT(10, 0). The phonon thermal conductance V.S. temperature, with fixed chemical potential.}
  \label{fig_t_z_sgm_ph_T}
\end{figure}
 so the electronic conductance is mainly contributed by the doping electrons in low temperature region. Due to the same reason, the EPI effect is more obvious in the SWCNT(10,0). For example, in panel (c), the reduction of $G_{e}$ can be as much as $50\%$ by the EPI, which is quite larger than that in the armchair tube.

Fig.~\ref{fig_t_z_Ge_u} shows the importance of temperature and chemical potential in the SWCNT(10,0). In low chemical potential region, the EPI is not so important; while the it becomes very important at high chemical potential. At 1000K, the $G_{e}$ is reduced by almost one half.

Fig.~\ref{fig_t_z_sgm_el_T} is for the electron thermal conductance versus temperature at different chemical potential. This figure is more or less similar as Fig.~\ref{fig_t_z_Ge_T} for
\begin{figure}[htpb]
  \begin{center}
    \scalebox{1.0}[1.0]{\includegraphics[width=7cm]{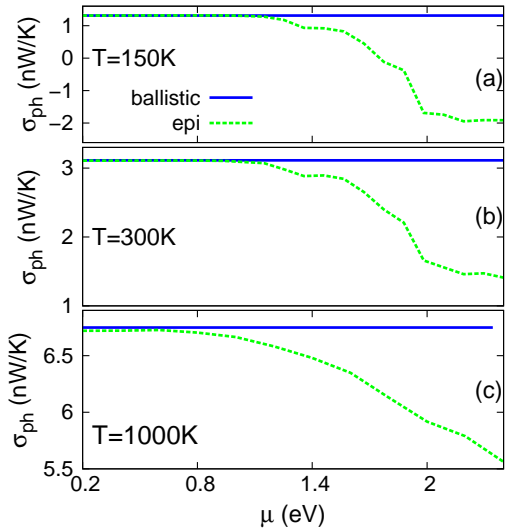}}
  \end{center}
  \caption{(Color online) SWCNT(10, 0). The the phonon thermal conductance V.S. chemical potential, at constant temperatures.}
  \label{fig_t_z_sgm_ph_u}
\end{figure}
 the electronic conductance, since the thermal current and electron charge current are carried by the same carriers (electron). For all three considered chemical potential, the EPI has limited effect in low temperature region and becomes very important in high temperature region. This result brings some information that, the EPI is more sensitive to the temperature in semiconductor zigzag tubes than that in the armchair tubes.

The phonon thermal conductance versus temperature is shown in Fig.~\ref{fig_t_z_sgm_ph_T}. It is quite similar as the armchair tubes, because the difference between these two types of tubes is in the nonzero band gap of the zigzag SWCNT(10, 0). However this band gap has no direct effect on the phonon thermal transport. The `electron-drag' effect for the zigzag SWCNT(10, 0) can also be seen from Fig.~\ref{fig_t_z_sgm_ph_u}~(a) where $T=150$ K. After $\mu>1.6$ eV, the phonons transport from low temperature region into high temperature region, driven by the dragging force of electrons through electron phonon scattering.

\section{conclusion}
To conclude, the EPI effect in SWCNT is investigated by using the NEGF approach within the Born approximation. The Joule heat and thermoelectric properties in both armchair SWCNT(10, 10) and semiconductor zigzag SWCNT(10, 0) are comparatively studied. It was found that the Joule heat is more sensitive to the chemical potential (gate voltage) than the temperature. With higher chemical potential, the Joule heat can reach a very large value by applying small bias voltage. The breakdown bias voltage for armchair SWCNT(10, 10) is estimated to be about 2.5 V, which is quite comparable with the experimental value 2.6 V. The Joule heat in zigzag SWCNT(10, 0) is considerably smaller than the armchair SWCNT(10, 10), resulting from the nonzero band gap in the electron band structure.

For the thermoelectric physical quantities, it was found that the EPI plays an important role at higher temperature or higher chemical potential for the $G_{e}$ and $\sigma_{el}$. At high chemical potential, the $G_{e}$ decreases with increasing temperature, which is opposite of the ballistic transport. This is the result of very strong electron phonon scattering. The phonon thermal conductance exhibits an interesting `electron-drag' phenomenon in case of low temperature and high chemical potential. In this situation, the phonons can be dragged by electrons from low temperature region into high temperature region with the help of EPI.

\textbf{Acknowledgements} The work is supported by a Faculty Research Grant of R-144-000-257-112 of NUS.


\begin{thebibliography}{}
\bibitem{BethuneDS} D. S. Bethune, C. H. Kiang, M. S. de Vries, G. Gorman, R. Savoy, J. Vazquez, and R. Beyers, Nature (London) \textbf{363}, 605 (1993).

\bibitem{IijimaS} S. Iijima and T. Ichihashi, Nature (London) \textbf{363}, 603 (1993).

\bibitem{BaughmanRH} R. H. Baughman, A. A. Zakhidov, W. A. de Heer, Science \textbf{297}, 787 (2002).

\bibitem{AvourisP} P. Avouris, Z. H. Chen, V. Perebeinos, Nat. Nanotechnol. \textbf{2}, 605 (2007).

\bibitem{CharlierJC} J. C. Charlier, X. Blase, S. Roche, Rev. Mod. Phys. \textbf{79}, 677 (2007).

\bibitem{WhiteCT} C. T. White, D. H. Robertson, and J. W. Mintmire, Phys. Rev. B \textbf{47}, 5485 (1993).

\bibitem{NiuC} C. Niu, E. K. Sickel, R. Hoch, D. Moy, H. Tennent, Appl. Phys. Lett. \textbf{70}, 1480 (1997).

\bibitem{KongJ} J. Kong, N. R. Franklin, C. W. Zhou, M. G. Chapline, S. Peng, K. J. Cho, and H. J. Dai, Science \textbf{287}, 622 (2000).

\bibitem{JaveyA2002} A. Javey, H. Kim, M. Brink, Q. Wang, A. Ural, J. Guo, P. McIntyre, P. McEuen, M. Lundstrom, H. J. Dai, Nature Mater. \textbf{1}, 241 (2002).

\bibitem{JaveyA2003} A. Javey, J. Guo, Q. Wang, M. Lundstrom, H. J. Dai, Nature \textbf{424}, 654 (2003).

\bibitem{WuJ} J. Wu, M. Eastman, T. Gutu, M. Wyse, J. Jiao, S.-M. Kim, M. Mann, Y. Zhang, and K. B. K. Teo, Appl. Phys. Lett. \textbf{91}, 173122 (2007).

\bibitem{HuR} R. Hu, B. A. Cola, N. Haram, J. N. Barisci, S. Lee, S. Stoughton, G. Wallace, C. Too, M. Thomas, A. Gestos, M. E. dela Cruz, J. P. Ferraris, A. A. Zakhidov, and R. H. Baughman, Nano. Lett \textbf{10}, 838 (2010).

\bibitem{SeidelRV} R. V. Seidel, A. P. Graham, B. Rajasekharan, E. Unger, M. Liebau, G. S. Duesberg, F. Kreupl, and W. Hoenlein, J. Appli. Phys. \textbf{96}, 6694 (2004).

\bibitem{JaveyA2004} A. Javey, J. Guo, M. Paulsson, Q. Wang, D. Mann, M. Lundstrom, and H. Dai, Phys. Rev. Lett. \textbf{92}, 106804 (2004).

\bibitem{HataK} K. Hata, D. N. Futaba, K. Mizuno, T. Namai, M. Yumura, S. Iijima, Science \textbf{306}, 1362 (2004).

\bibitem{PopE2005} E. Pop, D. Mann, J. Cao, Q. Wang, K. Goodson, and H. Dai, Phys. Rev. Lett. \textbf{95}, 155505 (2005).

\bibitem{TsutsuiM} M. Tsutsui, Y. Taninouchi, S. Kurokawa, and A. Sakai, J. Appli. Phys. \textbf{100}, 094302 (2006).

\bibitem{PopE2007} E. Pop, D. Mann, K. E. Goodson, and H. Dai, J. Appli. Phys. \textbf{101}, 093710 (2007).

\bibitem{MahanGD1997} G. D. Mahan, B. Sales, and J. Sharp, Physic Today \textbf{50}, 42 (1997).

\bibitem{ChenG} G. Chen, M. S. Dresselhaus, G. Dresselhaus, J. P. Fleurial, and T. Caillat, Int. Mater. Rev. \textbf{48}, 45 (2003).

\bibitem{NolasGS} G. S. Nolas, J. Poon, and M. G. Kanatzidis, MRS Bull. \textbf{31}, 199 (2006).

\bibitem{DresselhausM} M. Dresselhaus, G. Chen, M. Y. Tang, R. G. Yang, H. Lee, D. Z. Wang, Z. F. Ren, J. P. Fleurial, and P. Gogna, Adv. Mater. (Weinheim, Ger.) \textbf{19}, 1043 (2007).

\bibitem{HochbaumAI} A. I. Hochbaum, R. Chen, R. D. Delgado, W. Liang, E. C. Garnett, M. Najarian, A. Majumdar, and P. Yang, Nature \textbf{451}, 163 (2008).

\bibitem{VoTTM} T. T. M. Vo, A. J. Williamson, V. Lordi, and G. Galli, Nano. Lett. \textbf{8}, 1111 (2008).

\bibitem{ChenX} X. Chen, Y. Wang, Y. Ma, T. Cui, and G. Zou, J. Phys. Chem. C \textbf{113}, 14001 (2009).

\bibitem{MarkussenTPRB} T. Markussen, A. P. Jauho, and M. Brandbyge, Phys. Rev. B \textbf{79}, 035415 (2009).

\bibitem{MartinP} P. Martin, Z. Aksamija, E. Pop, and U. Ravaioli, Phys. Rev. Lett. \textbf{102}, 125503 (2009).

\bibitem{DonadioD} D. Donadio and G. Galli, Phys. Rev. Lett. \textbf{102}, 195901 (2009).

\bibitem{MarkussenT} T. Markussen, A. P. Jauho, and M. Brandbyge, Phys. Rev. Lett. \textbf{103}, 055502 (2009).

\bibitem{NiX} X. Ni, G. Liang, J.-S. Wang, and B. Li, Appl. Phys. Lett. \textbf{95}, 192114 (2009).

\bibitem{JiangJW} J. W. Jiang, J.-S. Wang, and B. Li, arXiv(unpublish).

\bibitem{OhtaH} H. Ohta, S. Kim, Y. Mune, T. Mizoguchi, K. Nomura, S. Ohta, T. Nomura, Y. Nakanishi, Y. Ikuhara, M. Hirano, H. Hosono, K. Koumoto, Nature Mater. \textbf{6}, 129 (2007).

\bibitem{ChoiWS} W. S. Choi, H. Ohta, S. J. Moon, Y. S. Lee, and T. W. Noh, arXiv:0906.5391.

\bibitem{Brenner} D. W. Brenner , O. A. Shenderova , J. A. Harrison, S. J. Stuart, B. Ni, and S. B. Sinnott, J. Phys.:Condens. Matter \textbf{14}, 783 (2002).

\bibitem{Gale} J. D. Gale, JCS Faraday Trans., \textbf{93}, 629 (1997).

\bibitem{MahanGD2003} G. D. Mahan, Phys. Rev. B \textbf{68}, 125409 (2003).

\bibitem{HaugH} H. Haug and A.-P. Jauho, \textit{Quantum Kinetics in Transport and Optics of Semiconductors}, (Springer, Berlin, 1996).

\bibitem{LuJT} J. T. L$\ddot{u}$ and J.-S. Wang, Phys. Rev. B \textbf{76}, 165418 (2007).

\bibitem{WangJS2008} J.-S. Wang, J. Wang, and J.T. Lu, Eur. Phys. J. B \textbf{62}, 381 (2008).

\bibitem{ViljasJK} J. K. Viljas, J. C. Cuevas, F. Pauly, and M. Hafner, Phys. Rev, B \textbf{72}, 245415 (2005).

\bibitem{PaulssonM} M. Paulsson, T. Frederiksen, and M. Brandbyge, Phys. Rev. B \textbf{72}, 201101 (2005); \textbf{75}, 129901 (E) (2007).

\bibitem{FrederiksenT} T. Frederiksen, M. Paulsson, M. Brandbyge, and A. P. Jauho, Phys. Rev. B \textbf{75}, 205413 (2007).

\bibitem{Zeng} N. Zeng, Ph.D thesis, National Univ. Singapore (2008). At Http://staff.science.nus.edu.sg/~phywjs/NEGF/negf.html.

\bibitem{ChiangIW1} I. W. Chiang, B. E. Brinson, R. E. Smalley, J. L. Margrave, and R. H. Hauge, J. Phys. Chem. B \textbf{105}, 1157 (2001).

\bibitem{ChiangIW2} I. W. Chiang, B. E. Brinson, A. Y. Huang, P. A. Willis, M. J. Bronikowski, J. L. Margrave, R. E. Smalley, and R. H. Hauge, J. Phys. Chem. B \textbf{105}, 8297 (2001).

\bibitem{OuyangY} Y. OuYang and J. Guo, Appl. Phys. Lett. \textbf{94}, 263107 (2009).

\bibitem{ScarolaVW} V. W. Scarola and G. D. Mahan, Phys. Rev. B \textbf{66}, 205405 (2002).
\end{thebibliography}
\end{document}